\documentclass[traditabstract,letter]{aa}

\usepackage{graphicx}
\usepackage{natbib}

\begin{document}
\title{Amplitudes of solar-like oscillations: a new scaling relation}

   \subtitle{}

   \author{
Hans~Kjeldsen\inst{1} \and 
Timothy R. Bedding\inst{2}}

   \offprints{Tim Bedding}

   \institute{Department of Physics and Astronomy, Aarhus University, DK-8000 Aarhus C,
Denmark\\
\email{hans@phys.au.dk}
\and 
Sydney Institute for Astronomy (SIfA), School of Physics, University of Sydney 2006,
Australia\\
\email{bedding@physics.usyd.edu.au}
}

   \date{To appear in A\&A Letters}

\newcommand{\eqref}[1]{Eq.~(\ref{#1})}
\newcommand{\eqsreftwo}[2]{Eqs.~(\ref{#1}) and~(\ref{#2})}
\newcommand{\eqsrefthree}[3]{Eqs.~(\ref{#1}), (\ref{#2}) and~(\ref{#3})}
\newcommand{\eqrangeref}[2]{Eqs.~(\ref{#1}) to~(\ref{#2})}
\newcommand{\half}{{\textstyle\frac{1}{2}}}
\newcommand{\Dnu}{\mbox{$\Delta \nu$}}
\newcommand{\GOLF}{{\em GOLF\/}}
\newcommand{\MOST}{{\em MOST\/}}
\newcommand{\SOHO}{{\em SOHO\/}}
\newcommand{\Teff}{\mbox{$T_{\rm eff}$}}
\newcommand{\VIRGO}{{\em VIRGO\/}}
\newcommand{\WIRE}{{\em WIRE\/}}
\newcommand{\acena}{\mbox{$\alpha$~Cen~A}}
\newcommand{\acenb}{\mbox{$\alpha$~Cen~B}}
\newcommand{\acen}{\mbox{$\alpha$~Cen}}
\newcommand{\bhyi}{\mbox{$\beta$~Hyi}}
\newcommand{\bvir}{\mbox{$\beta$~Vir}}
\newcommand{\tcet}{\mbox{$\tau$~Cet}}
\newcommand{\cms}{\mbox{cm\,s$^{-1}$}}
\newcommand{\comment}[1]{{\bf [#1]}}
\newcommand{\dpav}{\mbox{$\delta$~Pav}}
\newcommand{\eboo}{\mbox{$\eta$~Boo}}
\newcommand{\ms}{\mbox{m\,s$^{-1}$}}
\newcommand{\muHz}{\mbox{$\mu$Hz}}
\newcommand{\muara}{\mbox{$\mu$~Ara}}
\newcommand{\mynote}[1]{{\bf\it [#1]}}
\renewcommand{\mynote}[1]{\relax}

\newcommand{\nuind}{\mbox{$\nu$~Ind}}
\newcommand{\taucet}{\mbox{$\tau$~Cet}}
\newcommand{\HP}{\mbox{$H_P$}}
\newcommand{\Fcon}{\mbox{$F_{\rm con}$}}
\newcommand{\tgran}{\mbox{$\tau_{\rm gran}$}}
\newcommand{\tosc}{\mbox{$\tau_{\rm osc}$}}
\newcommand{\vosc}{\mbox{$v_{\rm osc}$}}
\renewcommand{\vosc}{\mbox{$A_{\rm vel}$}}
\newcommand{\ampint}{\left(\frac{\delta L}{L}\right)}
\renewcommand{\ampint}{\mbox{$A$}}

\newcommand{\nuac}{\mbox{$\nu_{\rm ac}$}}
\newcommand{\numax}{\mbox{$\nu_{\rm max}$}}
\newcommand{\kepler}{{\em Kepler}}
\newcommand{\corot}{{\em CoRoT}}
\newcommand{\new}[1]{{\bf #1}}
\renewcommand{\new}[1]{{\relax #1}}

\abstract{Solar-like oscillations are excited by near-surface convection
  and are being observed in growing numbers of stars using ground and
  space-based telescopes.  We have previously suggested an empirical
  scaling relation to predict their amplitudes (Kjeldsen and Bedding 1995).
  This relation has found widespread use but it predicts amplitudes in
  F-type stars that are higher than observed.  Here we present a new
  scaling relation that is based on the postulate that the power in
  velocity fluctuations due to p-mode oscillations scales with stellar
  parameters in the same way as the power in velocity fluctuations due to
  granulation.  The new relation includes a dependence on the damping rate
  via the mode lifetime and should be testable using observations from the
  \corot\ and \kepler\ missions.  We also suggest scaling relations for the
  properties of the background power due to granulation and argue that both
  these and the amplitude relations should be applicable to red giant
  stars.  }


   \maketitle

\section{Introduction}

When the first attempts to detect solar-like oscillations were being made,
we suggested an empirical scaling relation to predict their amplitudes
\citep[][hereafter Paper~I]{K+B95-paperI}.  This relation (and its various
modifications -- see Sec.~\ref{sec.original}) has found widespread use in
asteroseismology as observations of solar-like oscillations have
accumulated.  It has also been applied to cases in which p-mode
oscillations are regarded as noise, namely searches for exoplanets
\citep[e.g.,][]{OTTJ2008,NGW2009,BMB2010} and even for their moons
\citep{SSS2010}.

The steady flow of results over recent years from ground-based observations
\citep[e.g., reviews by][]{AChDC2008,B+K2008}, together with the flood of
data now arriving from the \corot\ and \kepler\ space missions
\citep[e.g.,][]{MBA2008,GBChD2010} makes it timely to revisit the amplitude
scaling relation.  That is the aim of this Letter.

\section{The empirical relation} \label{sec.original}

The empirical relation in Paper~I was based on theoretical models by
\citet{ChD+F83}, which were the only ones available at the time.  Based on
these models, we suggested that the velocity amplitudes of solar-like
oscillations scale from star to star according to
\begin{equation}
  \vosc \propto \frac{L}{M},  \label{eq.vel-propto-LM}
\end{equation}
where $L$ is the stellar luminosity and $M$ is the mass.  

Subsequently, model calculations seemed to show that a better
match might be obtained with 
\begin{equation}
  \vosc \propto \left(\frac{L}{M}\right)^s,
\end{equation}
with $s \approx 1.5$ \citep{HBChD99} or 
with $s \approx 0.7$ (\citealt{SGA2005,SBG2007,SGT2007}; see also
\citealt{Hou2006}).

Meanwhile, observations indicated that amplitudes of F-type stars fall below
both of these relations.  To account for
this, \citet{K+B2001} suggested a modified scaling relation that included
the effective temperature.  This was
\begin{equation}
  \vosc \propto \frac{1}{g} \propto \frac{L}{M \Teff^4}.
\end{equation}
However, this modification was not based on any physical grounds.  In
this Letter we use simple physical arguments to derive a revised scaling
relation for the amplitudes of solar-like oscillations.

\section{A revised approach}

Solar-like oscillations are excited by convection, which is also the
process responsible for granulation.  Our basic assumption is that the
power in velocity fluctuations due to p-mode oscillations scales with
stellar parameters in the same way as the power in velocity fluctuations
due to granulation.  The latter is a strong function of frequency and we
will evaluate it at $\numax$, the frequency at which the p-mode
oscillations are centred.  Our task, therefore, is to find a scaling
relation for granulation power in velocity at~$\numax$.

In doing this, we can take advantage of the established scaling relation
for \numax.  This is based on the suggestion by \citet{BGN91} that \numax\
might be expected to be a fixed fraction of the acoustic cutoff frequency
(see also Paper~I):
\begin{equation}
 \numax \propto \nuac \propto \frac{M \Teff^{3.5}}{L}. \label{eq.numax.nuac}
\end{equation}
This relation agrees quite well with observations \citep{B+K2003} and also
with model calculations \citep{CHA2008}, \new{and has recently been given a
theoretical basis by \citet{BGD2011}.}

\subsection{Granulation power in velocity}  \label{sec.gran}

To estimate the velocity fluctuations caused by granulation, we use the
standard model due to \citet{Har85} for the power density spectrum:
\begin{equation}
  P_{\rm vel}(\nu) = \frac{4\sigma_{\rm vel}^2 \tgran}{1 + (2\pi\nu\tgran)^2}.  \label{eq.Pgran.vel}
\end{equation}
Here, $\sigma_{\rm vel}$ is the rms of the velocity fluctuations due to
granulation and \tgran\ specifies the granulation timescale.  Similar
components at lower frequencies arising from mesogranulation,
supergranulation and active regions are not relevant here.  \new{Towards}
low frequencies the power density spectrum described by
\eqref{eq.Pgran.vel} \new{becomes} flat and \new{approaches} a value of
$4\sigma_{\rm vel}^2 \tgran$.  \new{Towards higher frequencies} it drops to
half power at frequency~$(2\pi\tgran)^{-1}$, and falls off as a power law
with a slope of~$-2$.

\subsubsection{A relation for $\tgran$}

To derive a scaling relation for the granulation timescale as a function of
stellar parameters, we assume that the vertical speed of the convection
cells is proportional to the sound speed,~$c_s$, and that the vertical
distance travelled scales with the pressure scale height,~\HP.  Hence we
have
\begin{equation}
  \tgran \propto \frac{\HP}{c_s}. \label{eq.tgran}
\end{equation}
As argued in Paper~I (Sec.~2.1), \new{the ideal gas law implies that} the
sound speed near the surface scales approximately as
\begin{equation}
  c_s \propto \sqrt{\Teff}.  \label{eq.cs}
\end{equation}
Meanwhile, the pressure scale height (Sec.~3.2 of Paper~I, see also
\citealt{K+W90}) scales approximately as
\begin{equation}
  \HP \propto \frac{\Teff}{g} \propto \frac{L}{M \Teff^3}. \label{eq.HP}
\end{equation}
We therefore have
\begin{equation}
  \tgran \propto \frac{L}{M \Teff^{3.5}}.
\end{equation}
Comparing with \eqref{eq.numax.nuac} shows that the granulation timescale
scales inversely with the acoustic cutoff frequency:
\begin{equation}
  \tgran \propto \frac{1}{\nuac} \propto \frac{1}{\numax},  \label{eq.tgran.nuac}
\end{equation}
a relation that we have already used for pipeline-processing of data from
the \kepler\ mission \citep{HSB2009}.

\subsubsection{A relation for $P_{\rm vel}(\numax)$}

To evaluate \eqref{eq.Pgran.vel} at \numax, we note that $\numax \gg
1/\tgran$ in the Sun and so the proportionalities in
\eqsreftwo{eq.numax.nuac}{eq.tgran.nuac} ensure this will remain true for
other stars.  We can therefore replace the denominator in
\eqref{eq.Pgran.vel} by 1 to get
\begin{equation}
 P_{\rm vel}(\numax) \propto \sigma_{\rm vel}^2\tgran 
                     \propto \frac{\sigma_{\rm vel}^2}{\numax}. \label{eq.P.vel.numax}
\end{equation}

For observations of an unresolved star, the velocity fluctuations arise
from a large number of granules on its surface \new{(assumed to behave in a
statistically independent manner)} and we therefore expect the rms of these
fluctuations to scale inversely with the square root of the number of
granules.  We also assume that the vertical speed of the granules is
proportional to the sound speed,~$c_s$.  Therefore, the rms variation in
velocity due to granulation should scale as
\begin{equation}
  \sigma_{\rm vel} \propto \frac{c_s}{\sqrt{n}}.  \label{eq.sigma.vel}
\end{equation}

To estimate $n$, we assume the diameter of the granules to be proportional to
the pressure scale height of the atmosphere \citep{Sch75,ACN84,FHS97}.  In
this case, the number occupying the surface of a star scales as
\begin{equation}
  n \propto \left(\frac{R}{\HP}\right)^2,  \label{eq.n}
\end{equation}
where $R$ is the stellar radius.  

Combining \eqsreftwo{eq.sigma.vel}{eq.n}, we therefore have
\begin{equation}
  \sigma_{\rm vel} \propto  \frac{c_s \HP}{R} \propto \frac{L^{0.5}}{M \Teff^{0.5}},
  \label{eq.sigma.vel.final} 
\end{equation}
where we have eliminated $R$ using $L \propto R^2 \Teff^4$ and eliminated
$\HP$ using \eqref{eq.HP}.

Using \eqsreftwo{eq.numax.nuac}{eq.sigma.vel.final}, we can write
\eqref{eq.P.vel.numax} as
\begin{eqnarray}
  P_{\rm vel}(\numax) & \propto & \frac{L^2}{M^3
  \Teff^{4.5}}. \label{eq.P.vel.numax.final}
\end{eqnarray}
This is the scaling relation for granulation power in velocity at~\numax.

\subsection{Oscillation amplitudes in velocity}  \label{sec.amp}

The amplitudes of solar-like oscillations depend on both the excitation and
the damping rate, where the latter is given by the mode lifetime.  We
postulate that the squared amplitude of p-mode oscillations in velocity is
proportional to the mode lifetime multiplied by the velocity power density
of the granulation at $\numax$:
\begin{equation}
  \vosc^2 \propto P_{\rm vel}(\numax)\, \tosc.
\end{equation}
This gives
\begin{equation}
  \vosc \propto \frac{L \tosc^{0.5}}{M^{1.5}  \Teff^{2.25}}, \label{eq.vosc-new}
\end{equation}
which is our new scaling relation for velocity amplitudes to replace
\eqref{eq.vel-propto-LM}.  It now includes a strong dependence on \Teff,
and also a weak dependence on mode lifetime.  Stars with longer mode
lifetimes will show larger amplitudes, \new{all other things being equal}.
Efforts to establish a scaling relation for \tosc\ have been made
\citep{CHK2009,BBB2011} and the wealth of new data from \corot\ and
\kepler\ will hopefully confirm one soon.


\section{Comparison with observations}  \label{sec.obs}

The first point to note is that correlated variations in mode amplitude and
lifetime are seen in the Sun over the solar cycle.  \citet{CEI2000}
analysed velocity observations of the p-mode oscillations in the Sun over
the declining phases of activity cycle~22 and found a 24\% increase in mode
linewidths (that is, a decrease in \tosc) that was matched by an identical
decrease in modal velocity powers (that is, in $\vosc^2$).  This agrees
exactly with expectations from \eqref{eq.vosc-new} \new{and with the
inferences of \citet{CEI2000}, who showed (by analogy with a damped,
stochastically driven oscillator) that the relative sizes observed in
changes of the mode lifetimes and amplitudes could be explained by changes
in the damping alone (i.e., without the need to alter the net forcing of
the modes over the solar cycle).}
\new{This also leads us to suggest that variations in oscillation
amplitudes in other stars during their activity cycles, as recently
reported by \citet{GMS2010} for the star HD~49933 from CoRoT observations,
would also be due to changes in the damping.  That is, the input power
stays constant but the mode lifetime changes, and so the amplitude changes
according to \eqref{eq.vosc-new}.}

We now compare \eqref{eq.vosc-new} with observations of solar-like
oscillations.  We are limited to those stars with measured velocity
amplitudes and for which reliable mode lifetimes have been determined.
These are
\acena\ and~B \citep{KBB2005},
\tcet\ \citep{TKB2009},
Procyon\ \citep{BKC2010},
and \bhyi\ \citep{BKA2007}.  
Observed amplitudes were estimated using
the method described by \citet{KBA2008}, which involves smoothing the power
spectrum.  

\begin{figure*}
\resizebox{0.48\hsize}{!}{
\includegraphics{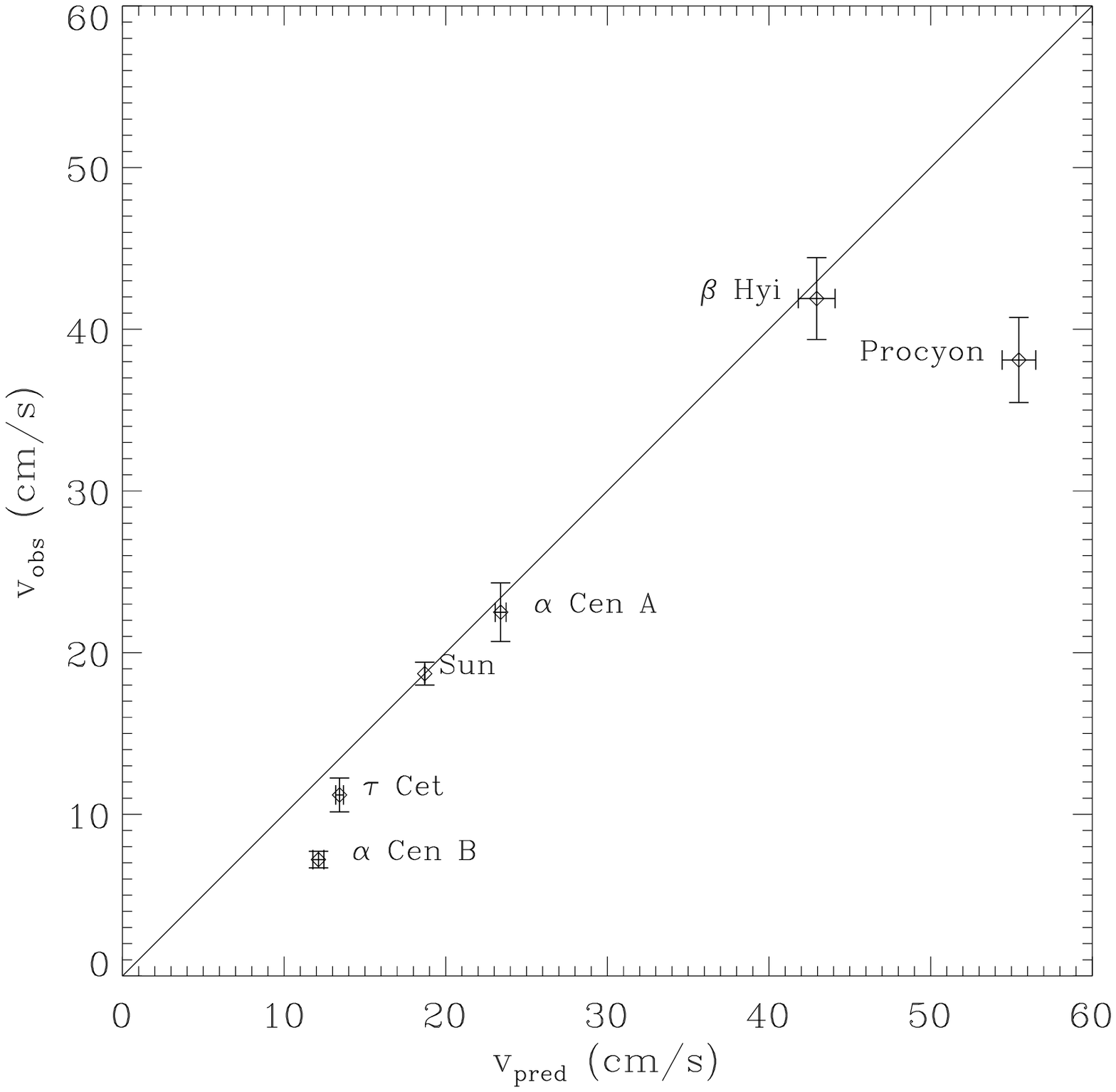}}
\hfill
\resizebox{0.48\hsize}{!}{
\includegraphics{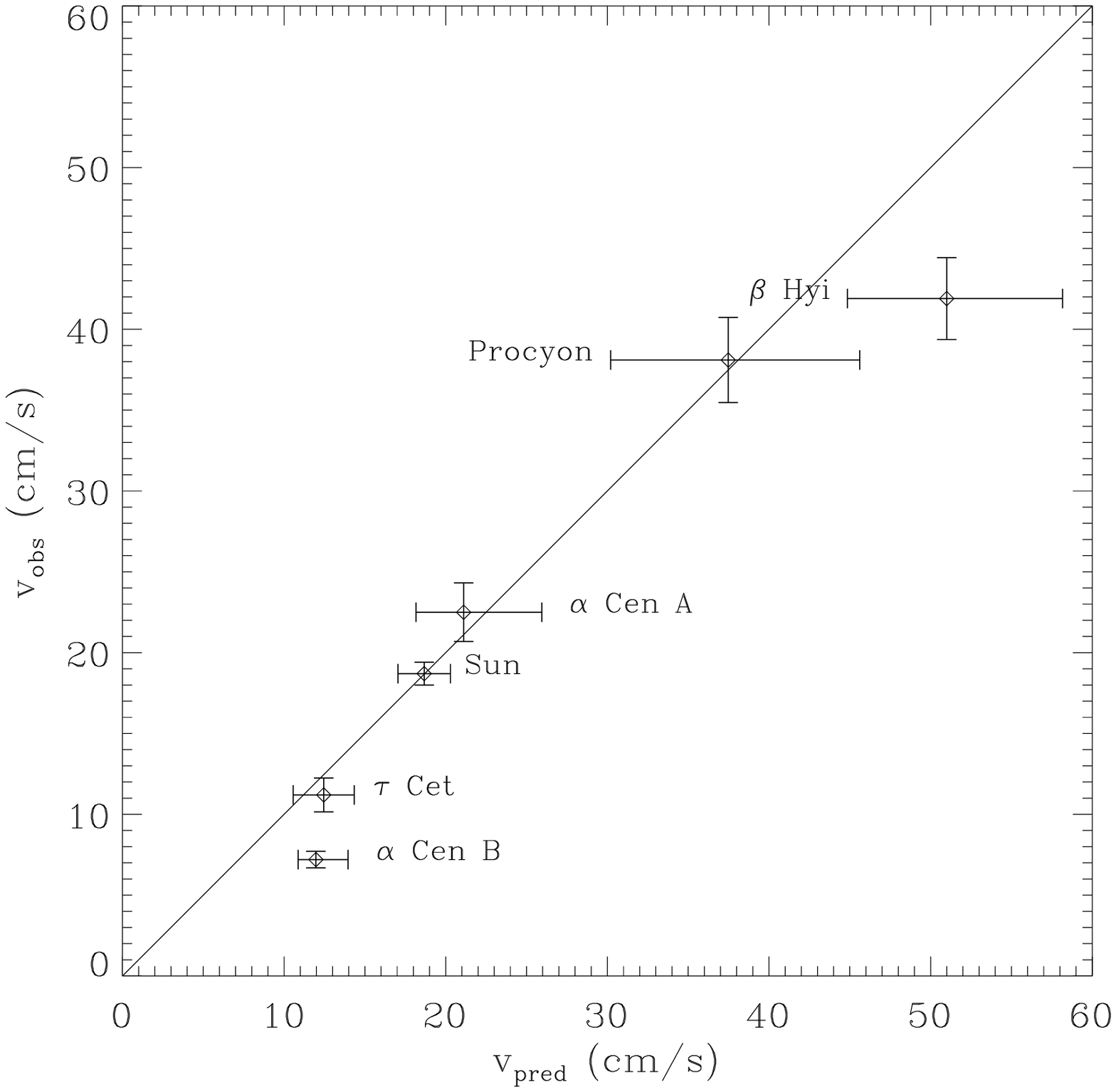}}
\caption{\label{fig.amp-lm} Velocity amplitudes of solar-like oscillations
  compared to the $(L/M)^{0.7}$ scaling (left) \label{fig.amp} and also
  compared to the new relation given by \eqref{eq.vosc-new} (right). }
\end{figure*}

To compare with other stars, we need a value for the mode lifetime in the
Sun at maximum power.  To establish this, we averaged the linewidths for
the 5 central values in Table 1 of \citet{CEI2001}, giving $\tosc_{,\odot}
= 2.88$\,d.  We then used \eqref{eq.vosc-new} to calculate expected
velocity amplitudes for those stars with published measured mode lifetimes.
The results are shown in Figure~\ref{fig.amp}.  \new{Most of these stars
have similar effective temperatures to the Sun and there is little to
choose between the two relations.  However, for the F-type star Procyon
($\Teff = 6500\,$K), which probes the domain where the old relation is
known to fail, the new relation gives much better agreement.}  Clearly more
stars with measured mode lifetimes are needed to confirm the result.
Detailed tests should soon be possible using the large number of stars now
being observed by the \corot\ and \kepler\ space missions.  However, those
missions are performing photometry and so we need to discuss amplitudes
measured in intensity.

\section{Intensity measurements}  \label{sec.photom}

\subsection{Oscillation amplitudes in intensity} \label{sec.osc.int}

The amplitudes of oscillations in velocity are directly related to the
velocity fluctuations from granulation, because the physical motion of
convective cells is what drives the oscillations.  The same is not true
when the oscillations are observed in intensity.  The intensity variations
arise primarily from temperature variations that are caused by the
compression and expansion of the atmosphere during the oscillation cycle.

In Paper~I we argued that for an adiabatic sound wave, the fractional
change in bolometric luminosity (integrated over all wavelengths) is
related to the velocity amplitude:
\begin{equation}
  \ampint_{\rm bol} \propto \frac{\vosc}{\sqrt{\Teff}}.
\end{equation}
In practice, observations are made in a certain wavelength range.  The
intensity amplitude observed at a wavelength $\lambda$, assuming this
wavelength is reasonably close to the peak of the blackbody spectrum (see
Sec.~2.2 in Paper~I), is
\begin{equation}
  \ampint_{\lambda} \propto \frac{\vosc}{\lambda
  \Teff^{r}} \label{eq.int-lambda}
\end{equation}
with $r=1.5$ (adopting the notation of \citealt{HBS2010}).  In fact, a fit
to observations of pulsating stars indicated $r=2.0$ (Fig.~1 in Paper~I).
This difference presumably reflects that fact that stellar
\new{oscillations} are not adiabatic.  We note that some authors have
chosen to adopt $r=1.5$ \citep[e.g.,][]{MBA2008,MBG2010}.

Combining \eqsreftwo{eq.vosc-new}{eq.int-lambda} gives our new scaling
relation for intensity amplitudes:
\begin{equation}
  \ampint_{\lambda} \propto \frac{L
  \tosc^{0.5}}{\lambda M^{1.5} \Teff^{2.25+r}},
\end{equation}
which can be tested with observations from \corot\ and \kepler.

\subsection{Granulation power in intensity}  \label{sec.gran.photom}

For completeness, we also make some comments about the intensity
fluctuations caused by granulation.  These arise because of surface
brightness variations from the contrast between cool and dark regions
\citep[e.g.,][]{TChDN98,S+L2005,Lud2006,LSS2009}.  We can estimate the
granulation power in intensity by again using the \citet{Har85} model:
\begin{equation}
  P_{\rm int}(\nu) = \frac{4\sigma_{\rm int}^2 \tgran}{1 +
  (2\pi\nu\tgran)^2}.  \label{eq.Pgran.int}
\end{equation}
The granulation timescale, \tgran, is the same as before, but we need a
scaling relation for the rms of the intensity fluctuations,~$\sigma_{\rm
int}$.  This should depend on the total number of granules that are visible
on the surface:
\begin{equation}
  \sigma_{\rm int} \propto \frac{1}{\sqrt{n}}. \label{eq.sigma.int}
\end{equation}
Choosing once again to measure the power at \numax, we can use the same
arguments as before (see \eqref{eq.P.vel.numax}) to get
\begin{equation}
 P_{\rm int}(\numax) \propto \frac{\sigma_{\rm int}^2}{\numax}. \label{eq.Pint.numax}
\end{equation}
Using \eqsrefthree{eq.HP}{eq.n}{eq.sigma.int}, we therefore have
\begin{equation}
 P_{\rm int}(\numax) \propto \frac{L^2}{M^3 \Teff^{5.5}}. \label{eq.Pgran.int.final}
\end{equation}
This scaling relation for the granulation background can be tested and
refined with data from \corot\ and \kepler.  However, we should note that
the intensity fluctuations caused by granulation depend on the contrast
between cool and dark regions, which in turn depends on the opacities and
the amount of limb darkening \citep{FHS97,S+L2005,Lud2006}.  We might therefore
expect some additional dependence on \Teff\ and also on metallicity.

\section{Red giants}

The original scaling relations discussed in Sec.~\ref{sec.original} were
based on models of stars on or close to the main sequence.  However, they
have also been applied to observations of red giants
\citep[e.g.,][]{E+G96,Gil2008,ZRH2008,S+G2009,HBS2010,MBG2010,BBB2011},
despite the fact that model calculations of red giants show poor agreement
\citep{H+G2002}.  The new relations proposed in this Letter might
reasonably be expected to apply to red giants.  However, it should be kept
in mind that red giants show additional $l=1$ mixed modes that may affect
the total amount of measured oscillation power
\citep[e.g.,][]{DBS2009,HBS2010}.  In addition, the coolest red giants have
molecular absorption features in their spectra that are sensitive to
temperature and would produce larger intensity amplitudes than expected
under the black-body assumption (Sec.~\ref{sec.osc.int}).  For the case of
granulation background, on the other hand, the relations in
\eqsreftwo{eq.tgran.nuac}{eq.Pgran.int.final} appear to be confirmed by
data from \kepler\ (S. Mathur et al., in prep.).

\begin{acknowledgements}
This work has been supported by the Danish Natural Science Research Council
and the Australian Research Council.  We thank Dennis Stello, Alfio
Bonanno, Benoit Mosser and the anonymous referee for helpful comments.
\end{acknowledgements}

\end{document}